\documentclass[aps,prl,showpacs,showkeys,superscriptaddress,twocolumn]{revtex4-1}
\usepackage[colorlinks,linkcolor=blue,anchorcolor=blue,citecolor=blue,urlcolor=blue,breaklinks=true]{hyperref}
\usepackage{graphicx}
\usepackage{amsmath}
\usepackage{amssymb}
\usepackage{slashed}
\usepackage{latexsym}
\usepackage{epsfig}
\usepackage{amsbsy}
\usepackage{array}
\usepackage{amssymb}
\usepackage{setspace}
\usepackage{bm}
\usepackage{lipsum}
\usepackage{mathrsfs}
\usepackage{float}
\usepackage{color}

\begin{document}

\title{New self-consistent mean field approximation and its application in strong interaction phase transition}

\author{Fei Wang}
\affiliation{Department of physics, Nanjing University, Nanjing 210093, China}
\author{Yakun Cao}
\affiliation{Department of physics, Nanjing University, Nanjing 210093, China}
\author{Yonghui Xia}
\affiliation{Department of physics, Nanjing University, Nanjing 210093, China}
\affiliation{Nanjing Proton Source Research and Design Center, Nanjing 210093, China}
\author{Hongshi Zong}
\email{zonghs@nju.edu.cn}
\affiliation{Department of physics, Nanjing University, Nanjing 210093, China}
\affiliation{Nanjing Proton Source Research and Design Center, Nanjing 210093, China}
\affiliation{Joint Center for Particle, Nuclear Physics and Cosmology, Nanjing 210093, China}

\begin{abstract}
In this letter, taking the  Nambu--Jona--Lasinio model as an example, we propose a new self-consistent mean field approximation method by means of Fierz transformation. This new self-consistent mean field approximation introduces a new free parameter $\alpha$ to be determined by experiments and when $\alpha$ takes 0.5, it reduces to the mean field approximation that was commonly used in the past. Then based on this self-consistent mean field approximation, we study the influence of the undetermined parameter $\alpha$ on the phase diagram of the two-flavor strong interaction matter. It is found that the value of $\alpha$ plays an extremely important role in the study of strong interaction phase diagram. It not only changes the position of the phase transition point of strong interaction matter, but also affects the order of phase transition, for example, when $\alpha$ is greater than the critical value $\alpha_c = 0.71$, then the strong interaction matter phase diagram no longer exists critical end point. In addition, in the case of zero temperature and finite density, we also found that when $\alpha$ is greater than 1.044, the pseudo-critical chemical potential is about 4$\sim$5 times the saturation density of the nuclear matter, which agrees with the expected  results from the image of hadrons degree of freedom. The resulting equations of state of strong interaction matter at low temperatures and high densities will have an important impact on  the study of the mass radius relationship of neutron stars and  the merging process of binary neutron stars.

\end{abstract}

\maketitle


It is generally believed that with increasing temperature or baryon chemical potential, strong interaction matter will undergo a phase transition from the hadronic matter to the quark-gluon plasma (QGP), which is expected to appear in the ultrarelativistic heavy ion collisions. The basic theory describing strong interactions is quantum chromodynamics (QCD).
Drawing a phase diagram of strong interaction matter at finite temperature and finite density is an important goal of current high nuclear physics. Although Lattice Monte Carlo simulations has made significant progress in the study of finite temperature and low chemical potential, it encounters the so called  sign problem when dealing with large chemical potentials. Therefore, the calculations of large chemical potentials based on effective theories of QCD are irreplaceable nowadays.

At present, people usually describe the phase transition of strong interaction matter under the condition of large chemical potential from two different physical images. One is based on the quark-gluon  degree of freedom, and the other is from the hadron degree of freedom. Let us first look at the phase transition of  strong interaction matter in the case of large chemical potentials derived from  quark-gluon degree of freedom. In Ref. \cite{Halasz:1998}, based on a universal argument, it is pointed out that when the chemical potential $\mu$ is smaller than a critical value $\mu_0$, the quark-number density vanishes identically.
 Namely, $ \mu_0$ is a singularity which separates two the regions with different quark-number densities. The numerical value of the critical chemical potential $\mu_0$ in pure QCD 
(i.e., with electromagnetic interactions being switched off) is estimated to be $\mu_0=(\mbox{M}_N -16 \mbox{MeV})/\mbox{N}_c = 307$ MeV (where $\mbox{M}_N$ is the nucleon mass and $\mbox{N}_c = 3$ is the number of colors). this means that at temperature $T=0$, and when $\mu$ is greater than $\mu_0$, the baryon matter will be excited from the QCD vacuum, which is  a robust and model-independent prediction \cite{Halasz:1998}. 
Along with the continuous increase of the quark chemical potential, it is believed that when the chemical potential is as large as a critical value $\mu_c$, the strong interaction matter will undergo the so-called chiral restoration and deconfinement phase transition \cite{Stephanov:2007, Fukushima:2010}. 
Of course, since people can't use the Lattice QCD to deal with the large chemical potential problem at present,  the value of $\mu_c$ depends on the phenomenological QCD model we choose. For example, the $\mu_c$ predicted by NJL model， which is commonly used, is about $330 \sim 380$ MeV \cite{Asakawa:1989, Buballa:2005, Cui:2018}. This means that the chemical potential required to excite a baryon from the QCD vacuum to the chemical potential required for a phase transition of a strong interaction  matter does not exceed dozens of MeV, such a result is hard to believe that is reasonable. However, from the image of hadron degrees of freedom, it is generally believed that strong interaction phase transitions are only possible when the density of nuclear matter is at least greater than 4 times the saturation density of the nuclear matter. So what is the quark chemical potential corresponding to the saturation density of four times nuclear matter? To see this clearly, we use the Relativistic Mean Field (RMF) method \cite{Walecka:1974} to estimate the quark chemical potential corresponding to the saturation density of one to five times the nuclear matter, as shown in TABLE. \ref{Table}.

\begin{table}[H]
	\begin{center}
		\begin{tabular}{ccccccc} 
			\hline
			\cline{1-7}
			nuclear density &$\rho_0$&$2\rho_0$&$3\rho_0$&$4\rho_0$&$5\rho_0$&\\
			\hline
			quark chemical potential (\mbox{MeV})&$323$&$371$&$469$&$580$&$692$&\\
			\hline
		\end{tabular}
		\caption{The quark chemical potential and corresponding nuclear matter density, $\rho_0$ is the nuclear matter saturation density. Results are given by the method of RMF with model parameters NL3$\omega\rho$ from the hadron degrees of freedom \cite{Fortin:2016,Li:2018a,Li:2018b}.}\label{Table}
	\end{center}
\end{table}
It can be clearly seen from TABLE. \ref{Table} that the quark chemical potential corresponding to the saturation density of the quadruple nuclear material is about 580 MeV. That is to say, from the image of hadron degrees of freedom, in the case of zero temperature and finite chemical potential, the phase transition of strong interaction matter is likely to occur only when the quark chemical potential is at least greater than 580 MeV.  Obviously, the position of the phase change of the strong interaction matter obtained from the perspective of the image of the hadron degree of freedom is significantly different from that derived from the quark-gluon degree of freedom. Therefore, there are huge contradictions between the results drawn from the quark-gluon degrees of freedom and the expected results derived from the hadron degrees of freedom. The main motivation of this letter is to try to propose a new \emph{self-consistent} mean field approximation to solve the above contradictions.

Before a new approach of self-consistent mean field approximation is proposed, let us first review the mean field approximation that is usually adopted in the past.  Mean field approximation is to replace all interactions to any one body with an average or effective interaction, which  reduces any many body problem into an effective one-body problem. Taking the Nambu--Jona--Lasinio (NJL) model \cite{Nambu:1961a,Nambu:1961b} as an example and the Lagrangian is given by
\begin{equation}
	\mathscr{L}=\bar\psi (i\slashed{\partial}-m)\psi+G[ (\bar\psi\psi)^2+(\bar\psi i\gamma^5\cdot \tau \psi)^2], \label{Lagrangian1}
\end{equation}
where $m$ denotes current quark mass and $G$ denotes the coupling constant. Performing  Fierz transformation on the interaction part of the Lagrangian (\ref{Lagrangian1}) in   flavor, Dirac spinor and color spaces, one has
\begin{eqnarray}
	&&\mathscr{F}[(\bar\psi\psi)^2+(\bar\psi i\gamma^5\cdot \tau \psi)^2]= \nonumber \\
	&&\frac{1}{8N_c}[2(\bar\psi\psi)^2+2(\bar\psi i \gamma^5\tau\psi)^2-2(\psi\tau\psi)^2-2(\bar\psi i \gamma^5\psi)^2-\nonumber\\
	&&4(\bar\psi\gamma^{\mu}\psi)^2-4(\bar\psi i\gamma^{\mu}\gamma^{5}\psi)^2+(\bar\psi\sigma^{\mu\nu}\psi)^2-(\bar\psi\sigma^{\mu\nu}\tau\psi)^2],
\end{eqnarray}
and the Lagrangian becomes 
\begin{eqnarray}
	\mathscr{L}_{F}=\bar\psi (i\slashed{\partial}-m)\psi+\mathscr{F}[(\bar\psi\psi)^2+(\bar\psi i\gamma^5\cdot \tau \psi)^2].
\end{eqnarray}
Because Fierz transformation is a mathematical equivalent transformation, therefore, the original Lagrangian $\mathscr{L}$ and the Fierz-transformed Lagrangian  $\mathscr{L}_{F}$  are equivalent. But here we need to point out that when we use the mean field approximation, and especially in the case of an external field, $\langle\mathscr{L}\rangle_{m}$ is no longer equal to $\langle \mathscr{L}_{F}\rangle_{m}$ (notation $\langle \cdots \rangle_{m}$ denotes mean field approximation). For example, when we study the strong interaction  phase transition  at finite chemical potential, which can  be regarded as the vector background field, the position of the phase transition calculated by $\langle\mathscr{L}\rangle_{m}$ and $\langle\mathscr{L}_{F}\rangle_{m}$ is very different \cite{Klevansky:1992}. Then, a question arises: which Lagrangian in mean field approximation is suitable? The Ref. \cite{Klevansky:1992} suggests the form $1/2(\langle\mathscr{L}\rangle_{m} +\langle\mathscr{L}_{F}\rangle_{m})$, because the form is Fierz invariant and both the Hartree term and Fock term contributions are included. Formally, as the Hartree term and  the Fock term contributions are identical, $\langle\mathscr{L}\rangle_{m}$ and  $\langle \mathscr{L}_{F}\rangle_{m}$ contributions are equal. In fact, there is not any physical requirements for that, the  Hartree term and  the Fock term contributions are identical.

 Considering that the  original Lagrangian and  Fierz-transformed Lagrangian are identical, the most general effective Lagrangian can be redefined  as $\mathscr{L}_{R}=(1-\alpha)\mathscr{L}+\alpha\mathscr{L}_{F}$, in which $\alpha$ is an arbitrary \emph{c} number. Obviously $\mathscr{L}_{R}=\mathscr{L}$, that is to say, by redefining the effective  Lagrangian of the system, it does not change the original Lagrangian of the system. However, in the presence of chemical potential, once the mean field approximation is made, the situation will change greatly (see below). Performing  mean field approximation on the redefined Lagrangian, one has 
 \begin{eqnarray}
 \langle \mathscr{L}_{R}\rangle_m= (1-\alpha)\langle\mathscr{L}\rangle_m+\alpha\langle\mathscr{L}_{F}\rangle_m. \label{Lagrangianr}
   \end{eqnarray}
 
 The gap equation is then given by
 \begin{eqnarray}
 	M&&=m+\nonumber\\
 	&&\left( 24- 22 \alpha  \right) G\frac { M } { \pi ^ { 2 } } \int^{\Lambda} \frac { p ^ { 2 } } { E _ { p } } [ 1 - n _ { p } \left( T , \mu _ { r } \right) - \overline { n } _ { p } \left( T , \mu _ { r } \right)]\mbox{d} p, \nonumber\\ \label{gap}
 \end{eqnarray}
 where
 \begin{eqnarray}
 	\mu_{r}&&=\mu-\nonumber\\
 	&&\frac {  \alpha G } { N _ { c } }\frac { 12 } { \pi ^ { 2 } } \int^{\Lambda} p ^ { 2 }  [  n _ { p } \left( T , \mu _ { r } \right) - \overline { n } _ { p } \left( T , \mu _ { r } \right)]  \mbox{d} p. \label{mur}
 \end{eqnarray}
 Here, $E_{p}=\sqrt{p^2+M^2}$ and 
 \begin{eqnarray}
 	\begin{aligned} n _ { p } ( T , \mu_{r} ) & = \frac { 1 } { 1 + \exp \left( \frac { E _ { p } - \mu _ { r } } { T } \right) }, \\ \overline { n } _ { p } ( T , \mu_{r} ) & = \frac { 1 } { 1 + \exp \left( \frac { E _ { p } + \mu _ { r } } { T } \right) },\end{aligned}
 \end{eqnarray}
 where the parameter $\alpha$ can not be determined in advance by theory, but can only be determined by fitting experimental results. As mentioned above, in the case of finite chemical potential, the new self-consistent mean field approximation is quite different from the previous mean field approximation. When the chemical potential approaches zero, the gap equation obtained from the  self-consistent mean field approximation is reduced to the gap equation obtained from the conventional mean field approximation. At the same time, since the NJL model parameters are obtained by fitting the pion mass, decay constant and chiral condensation at zero temperature and zero chemical potential, the model parameters of this letter are the same as the previous NJL model. In addition,  we also stress that the form $1/2(\langle\mathscr{L}\rangle_{m} +\langle\mathscr{L}_{F}\rangle_{m})$  proposed by Ref. \cite{Klevansky:1992} is only a special case with $\alpha=1/2$. Similarly, $\alpha=0$, then $\langle \mathscr{L}_{R}\rangle_{m}=\langle\mathscr{L}\rangle_{m}$, only the Hartree term is under considering while the only the Fock term is under considering with $\alpha=1$, $\langle \mathscr{L}_{R}\rangle_{m} =\langle\mathscr{L}_{F}\rangle_{m}$. Here, it should be noted  that the effective Lagrangian defined by Eq. (\ref{Lagrangianr}) is  universal and can be applied in any strongly interacting system. For example, an important challenge of modern condensed matter physics is to develop a self-consistent method to determine the leading and sub-leading phase transition instabilities caused by various kinds of interaction. Recently this issue has been widely studied in iron-based superconductors \cite{Chubukov:2016} and Dirac semimetal materials \cite{Honerkamp:2008,Roy:2018,Tang:2018}, but many problems have not been satisfactorily resolved. In these strong correlated systems, the fermionic excitations are subjected to several sorts of interactions, which lead to a number of competing long-range orders. The approach developed in this letter could be applied to study the phase-transition instabilities and ordering competition in condensed matter physics.

 Due to the lack of experimental data at large chemical potential,  the parameter $\alpha$ can not be determined by experimental results and can be regarded as a free parameter. 
 In this letter, we first study that the chiral  phase transition depends on the parameter $\alpha$ at finite chemical potential and zero temperature and investigate the $\alpha$ effects on chiral phase transition. To determine the order of phase transition and the location of it,  the chiral susceptibility \cite{Cui:2015}
\begin{eqnarray}
		\chi_{m}=-\frac{\partial\langle\bar\psi\psi\rangle}{\partial m},
\end{eqnarray}
with different  $\alpha$ is adopted. 

\begin{figure}[htp!]
{\includegraphics[width=0.8\columnwidth]{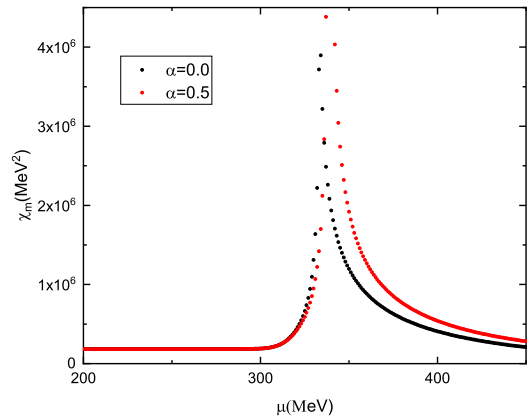} }
\caption{The chiral susceptibility as a function of chemical potential  at zero temperature ($\alpha$ is less than $\alpha_c=0.71$ ). The parameters of NJL model are given by  $G=4.93\times 10^{-6} \mbox{MeV}^{-2}$, $\Lambda=653 \mbox{MeV}$ and $m=5.0 \mbox{MeV}$ \cite{He:2005}.}
 \label{FIG1}
\end{figure}

\begin{figure}
{\includegraphics[width=0.8\columnwidth]{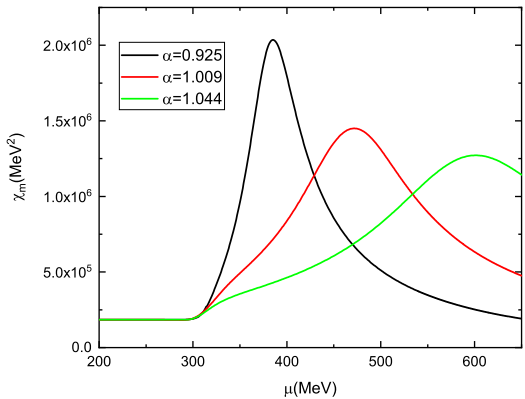} }
\caption{The chiral susceptibility as a function of chemical potential at zero temperature ( $\alpha$ is larger than $\alpha_c=0.71$ ). The parameters of NJL model are given as the same in Fig. \ref{FIG1}.}
 \label{FIG2}	
\end{figure}

Based on the numerical iterative algorithm, it is easy to numerically solve the quark gap equation at finite chemical potential and obtain the corresponding quark number density and chiral susceptibility, and we have the following interesting results: First, the critical value $\mu_{0}$ does not depend on the parameter $\alpha$, that is, no matter how much $\alpha$ is, the value of $\mu_0$ is always the same. For example, based on the model parameters adopted in this letter, $\mu_{0}$ is always equal to 311 $\mbox{MeV}$, which is quantitatively consistent with  the prediction of Ref. \cite{Halasz:1998}. Second, the chiral susceptibility shows different behaviors with different $\alpha$.  As $\alpha$ is less than the critical value $\alpha_c=0.71$, the chiral susceptibility is discontinuous at the critical chemical potential, therefore, the phase transition is a first order phase transition, see Fig. \ref{FIG1}. It should be pointed out here that, in the previous mean field approximation, $\alpha$ is often taken as 0.5, which is less than $\alpha_c = 0.71$, therefore the strong interaction phase transition at low temperature and large chemical potential predicted by NJL model \cite{Klevansky:1992} is the first order phase transition. Third, with the increases of $\alpha$, the chiral susceptibility shows a smooth peak at the pseudo-critical chemical potential and the phase transition is a crossover, see Fig. \ref{FIG2}. Also, the pseudo-critical chemical potential increases with the parameter $\alpha$ at zero temperature. Therefore, the parameter $\alpha$ not only affects the value of the (pseudo)critical chemical potential, but also affects the order of the phase transition. 
Just as mentioned above, the parameter $\alpha$ can not be determined by experiment currently, however, people generally believe that the phase transition of strong interaction matter is at least 4 times the saturation density of nuclear matter. If this is considered as a physical requirements to constrain $\alpha$, we find that when $\alpha=1.044$, that is, the corresponding  pseudo-critical chemical potential $\mu=600$ MeV, strong interaction matter phase transition  will occur. Obviously, the self-consistent mean field method described above can be easily extended from 2 flavors to $2+1$ flavors strong interaction matter, and thus the equation of state at zero temperature and finite chemical potential is obtained. This will provide a good basis for us to study the neutron star mass radius relationship \cite{Li:2018a,Li:2018b}.

\begin{figure}
{\includegraphics[width=0.8\columnwidth]{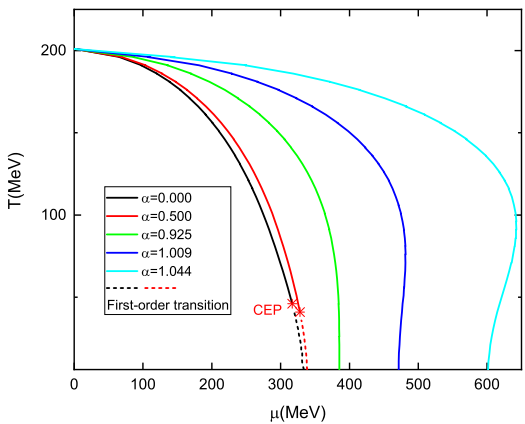} }
\caption{The QCD phase diagram with different parameter $\alpha$.}
 \label{FIG3}	
\end{figure}

Now let us extend the above approach from zero temperature to finite temperature.
At finite temperature with vanishing chemical potential, the chiral phase transition is a crossover, which is confirmed by lattice simulations\cite{Borsanyi:2010}. Since the fermion sign problem, lattice simulations can not perform calculations at large chemical potential.   Lots of models shows that the chiral phase transition at finite chemical potential is a first order phase transition \cite{Stephanov:2007,Cui:2015}. Thus, the first order phase transition must be terminated at the critical temperature and chemical potential, which is the critical end point (CEP). Locating the possible CEP is an important target in the second phase energy scanning plan of Relativistic Heavy Ion Collider (RHIC) \cite{Aggarwal:2010, Adamczyk:2014,Luo:2015, Luo:2017, Adamczyk:2017, Gupta:2011}. However, in the new consistent mean field approach, the existence of CEP depends on the parameter $\alpha$, see Fig. \ref{FIG3}. As indicated in  Fig. \ref{FIG3}, with the increases of $\alpha$, the CEP does not exist, i.e., if  $\alpha$ is larger than the critical value $\alpha_c=0.71$, the chiral phase transition is a crossover at finite temperature and chemical potential and no CEP  exists. Obviously, the above results are model dependent. Our results are valid only if the mean field approximation can capture the nature of the strong interaction matter phase transition.
    
In conclusion, we proposed a new approach of  \emph{self-consistent} mean field approximation, which is universal and we apply it in the study  of two flavor strong interaction matter phase transition. In the  new approach, a free parameter $\alpha$ is introduced, which can not be fixed in advance by theory. We study that the effects of $\alpha$ on chiral phase transition and find that the parameter not only affects the value of the (pseudo)critical chemical potential, but also affects the order of the phase transition. It is also found that the (pseudo)critical chemical potential increases with the parameter $\alpha$. When $\alpha = 1.044$, the pseudo-critical chemical potential is even about $600 \mbox{MeV}$ , which agrees with the expected results from the image of hadron degrees of freedom. Moreover, with the increasing of $\alpha$, the chiral phase transition becomes a crossover and therefore there is no CEP in QCD phase diagram at finite temperature and finite chemical potential. Finally, what we need to emphasize is that determining $\alpha$ is a very interesting question. It may be determined by the observation of  the binary neutron star merger, such as GW170817 \cite{Abbott:2017prl} and  the mass constraint of PSR  J$0348+0432$ \cite{Antoniadis:2013}. This is what we are going to investigate in the future.

The authors thank Professor Liu Guozhu and Luo Xiaofeng for their helpful discussions. This work is supported in part by the National Natural Science Foundation of
China (under Grants  No.11690030, No.11475085,  No.11535005 ) and National Major state Basic Research and Development of China (2016YFE0129300).

\bibliography{reference}

\begin{thebibliography}{28}%
\makeatletter
\providecommand \@ifxundefined [1]{%
 \@ifx{#1\undefined}
}%
\providecommand \@ifnum [1]{%
 \ifnum #1\expandafter \@firstoftwo
 \else \expandafter \@secondoftwo
 \fi
}%
\providecommand \@ifx [1]{%
 \ifx #1\expandafter \@firstoftwo
 \else \expandafter \@secondoftwo
 \fi
}%
\providecommand \natexlab [1]{#1}%
\providecommand \enquote  [1]{``#1''}%
\providecommand \bibnamefont  [1]{#1}%
\providecommand \bibfnamefont [1]{#1}%
\providecommand \citenamefont [1]{#1}%
\providecommand \href@noop [0]{\@secondoftwo}%
\providecommand \href [0]{\begingroup \@sanitize@url \@href}%
\providecommand \@href[1]{\@@startlink{#1}\@@href}%
\providecommand \@@href[1]{\endgroup#1\@@endlink}%
\providecommand \@sanitize@url [0]{\catcode `\\12\catcode `\$12\catcode
  `\&12\catcode `\#12\catcode `\^12\catcode `\_12\catcode `\%12\relax}%
\providecommand \@@startlink[1]{}%
\providecommand \@@endlink[0]{}%
\providecommand \url  [0]{\begingroup\@sanitize@url \@url }%
\providecommand \@url [1]{\endgroup\@href {#1}{\urlprefix }}%
\providecommand \urlprefix  [0]{URL }%
\providecommand \Eprint [0]{\href }%
\providecommand \doibase [0]{http://dx.doi.org/}%
\providecommand \selectlanguage [0]{\@gobble}%
\providecommand \bibinfo  [0]{\@secondoftwo}%
\providecommand \bibfield  [0]{\@secondoftwo}%
\providecommand \translation [1]{[#1]}%
\providecommand \BibitemOpen [0]{}%
\providecommand \bibitemStop [0]{}%
\providecommand \bibitemNoStop [0]{.\EOS\space}%
\providecommand \EOS [0]{\spacefactor3000\relax}%
\providecommand \BibitemShut  [1]{\csname bibitem#1\endcsname}%
\let\auto@bib@innerbib\@empty
\bibitem [{\citenamefont {Halasz}\ \emph {et~al.}(1998)\citenamefont {Halasz},
  \citenamefont {Jackson}, \citenamefont {Shrock}, \citenamefont {Stephanov},\
  and\ \citenamefont {Verbaarschot}}]{Halasz:1998}%
  \BibitemOpen
  \bibfield  {author} {\bibinfo {author} {\bibfnamefont {M.~A.}\ \bibnamefont
  {Halasz}}, \bibinfo {author} {\bibfnamefont {A.~D.}\ \bibnamefont {Jackson}},
  \bibinfo {author} {\bibfnamefont {R.~E.}\ \bibnamefont {Shrock}}, \bibinfo
  {author} {\bibfnamefont {M.~A.}\ \bibnamefont {Stephanov}}, \ and\ \bibinfo
  {author} {\bibfnamefont {J.~J.~M.}\ \bibnamefont {Verbaarschot}},\ }\href
  {\doibase 10.1103/PhysRevD.58.096007} {\bibfield  {journal} {\bibinfo
  {journal} {Phys. Rev.}\ }\textbf {\bibinfo {volume} {D58}},\ \bibinfo {pages}
  {096007} (\bibinfo {year} {1998})}\BibitemShut {NoStop}%
\bibitem [{\citenamefont {Stephanov}(2006)}]{Stephanov:2007}%
  \BibitemOpen
  \bibfield  {author} {\bibinfo {author} {\bibfnamefont {M.~A.}\ \bibnamefont
  {Stephanov}},\ }\href {\doibase 10.22323/1.032.0024} {\bibfield  {journal}
  {\bibinfo  {journal} {PoS}\ }\textbf {\bibinfo {volume} {LAT2006}},\ \bibinfo
  {pages} {024} (\bibinfo {year} {2006})}\BibitemShut {NoStop}%
\bibitem [{\citenamefont {Fukushima}\ and\ \citenamefont
  {Hatsuda}(2011)}]{Fukushima:2010}%
  \BibitemOpen
  \bibfield  {author} {\bibinfo {author} {\bibfnamefont {K.}~\bibnamefont
  {Fukushima}}\ and\ \bibinfo {author} {\bibfnamefont {T.}~\bibnamefont
  {Hatsuda}},\ }\href {http://stacks.iop.org/0034-4885/74/i=1/a=014001}
  {\bibfield  {journal} {\bibinfo  {journal} {Reports on Progress in Physics}\
  }\textbf {\bibinfo {volume} {74}},\ \bibinfo {pages} {014001} (\bibinfo
  {year} {2011})}\BibitemShut {NoStop}%
\bibitem [{\citenamefont {Asakawa}\ and\ \citenamefont
  {Yazaki}(1989)}]{Asakawa:1989}%
  \BibitemOpen
  \bibfield  {author} {\bibinfo {author} {\bibfnamefont {M.}~\bibnamefont
  {Asakawa}}\ and\ \bibinfo {author} {\bibfnamefont {K.}~\bibnamefont
  {Yazaki}},\ }\href {\doibase 10.1016/0375-9474(89)90002-X} {\bibfield
  {journal} {\bibinfo  {journal} {Nucl. Phys.}\ }\textbf {\bibinfo {volume}
  {A504}},\ \bibinfo {pages} {668} (\bibinfo {year} {1989})}\BibitemShut
  {NoStop}%
\bibitem [{\citenamefont {Buballa}(2005)}]{Buballa:2005}%
  \BibitemOpen
  \bibfield  {author} {\bibinfo {author} {\bibfnamefont {M.}~\bibnamefont
  {Buballa}},\ }\href {\doibase https://doi.org/10.1016/j.physrep.2004.11.004}
  {\bibfield  {journal} {\bibinfo  {journal} {Physics Reports}\ }\textbf
  {\bibinfo {volume} {407}},\ \bibinfo {pages} {205} (\bibinfo {year}
  {2005})}\BibitemShut {NoStop}%
\bibitem [{\citenamefont {Cui}\ \emph {et~al.}(2018)\citenamefont {Cui},
  \citenamefont {Xu}, \citenamefont {Li}, \citenamefont {Sun}, \citenamefont
  {Zhang},\ and\ \citenamefont {Zong}}]{Cui:2018}%
  \BibitemOpen
  \bibfield  {author} {\bibinfo {author} {\bibfnamefont {Z.-F.}\ \bibnamefont
  {Cui}}, \bibinfo {author} {\bibfnamefont {S.-S.}\ \bibnamefont {Xu}},
  \bibinfo {author} {\bibfnamefont {B.-L.}\ \bibnamefont {Li}}, \bibinfo
  {author} {\bibfnamefont {A.}~\bibnamefont {Sun}}, \bibinfo {author}
  {\bibfnamefont {J.-B.}\ \bibnamefont {Zhang}}, \ and\ \bibinfo {author}
  {\bibfnamefont {H.-S.}\ \bibnamefont {Zong}},\ }\href {\doibase
  10.1140/epjc/s10052-018-6264-4} {\bibfield  {journal} {\bibinfo  {journal}
  {Eur. Phys. J.}\ }\textbf {\bibinfo {volume} {C78}},\ \bibinfo {pages} {770}
  (\bibinfo {year} {2018})}\BibitemShut {NoStop}%
\bibitem [{\citenamefont {Walecka}(1974)}]{Walecka:1974}%
  \BibitemOpen
  \bibfield  {author} {\bibinfo {author} {\bibfnamefont {J.~D.}\ \bibnamefont
  {Walecka}},\ }\href {\doibase https://doi.org/10.1016/0003-4916(74)90208-5}
  {\bibfield  {journal} {\bibinfo  {journal} {Annals Phys.}\ }\textbf {\bibinfo
  {volume} {83}},\ \bibinfo {pages} {491} (\bibinfo {year} {1974})}\BibitemShut
  {NoStop}%
\bibitem [{\citenamefont {Fortin}\ \emph {et~al.}(2016)\citenamefont {Fortin},
  \citenamefont {Providencia}, \citenamefont {Raduta}, \citenamefont
  {Gulminelli}, \citenamefont {Zdunik}, \citenamefont {Haensel},\ and\
  \citenamefont {Bejger}}]{Fortin:2016}%
  \BibitemOpen
  \bibfield  {author} {\bibinfo {author} {\bibfnamefont {M.}~\bibnamefont
  {Fortin}}, \bibinfo {author} {\bibfnamefont {C.}~\bibnamefont {Providencia}},
  \bibinfo {author} {\bibfnamefont {A.~R.}\ \bibnamefont {Raduta}}, \bibinfo
  {author} {\bibfnamefont {F.}~\bibnamefont {Gulminelli}}, \bibinfo {author}
  {\bibfnamefont {J.~L.}\ \bibnamefont {Zdunik}}, \bibinfo {author}
  {\bibfnamefont {P.}~\bibnamefont {Haensel}}, \ and\ \bibinfo {author}
  {\bibfnamefont {M.}~\bibnamefont {Bejger}},\ }\href {\doibase
  10.1103/PhysRevC.94.035804} {\bibfield  {journal} {\bibinfo  {journal} {Phys.
  Rev.}\ }\textbf {\bibinfo {volume} {C94}},\ \bibinfo {pages} {035804}
  (\bibinfo {year} {2016})}\BibitemShut {NoStop}%
\bibitem [{\citenamefont {Li}\ \emph {et~al.}(2018{\natexlab{a}})\citenamefont
  {Li}, \citenamefont {Yan}, \citenamefont {Geng}, \citenamefont {Huang},\ and\
  \citenamefont {Zong}}]{Li:2018a}%
  \BibitemOpen
  \bibfield  {author} {\bibinfo {author} {\bibfnamefont {C.-M.}\ \bibnamefont
  {Li}}, \bibinfo {author} {\bibfnamefont {Y.}~\bibnamefont {Yan}}, \bibinfo
  {author} {\bibfnamefont {J.-J.}\ \bibnamefont {Geng}}, \bibinfo {author}
  {\bibfnamefont {Y.-F.}\ \bibnamefont {Huang}}, \ and\ \bibinfo {author}
  {\bibfnamefont {H.-S.}\ \bibnamefont {Zong}},\ }\href {\doibase
  10.1103/PhysRevD.98.083013} {\bibfield  {journal} {\bibinfo  {journal} {Phys.
  Rev.}\ }\textbf {\bibinfo {volume} {D98}},\ \bibinfo {pages} {083013}
  (\bibinfo {year} {2018}{\natexlab{a}})}\BibitemShut {NoStop}%
\bibitem [{\citenamefont {Li}\ \emph {et~al.}(2018{\natexlab{b}})\citenamefont
  {Li}, \citenamefont {Zhang}, \citenamefont {Yan}, \citenamefont {Huang},\
  and\ \citenamefont {Zong}}]{Li:2018b}%
  \BibitemOpen
  \bibfield  {author} {\bibinfo {author} {\bibfnamefont {C.-M.}\ \bibnamefont
  {Li}}, \bibinfo {author} {\bibfnamefont {J.-L.}\ \bibnamefont {Zhang}},
  \bibinfo {author} {\bibfnamefont {Y.}~\bibnamefont {Yan}}, \bibinfo {author}
  {\bibfnamefont {Y.-F.}\ \bibnamefont {Huang}}, \ and\ \bibinfo {author}
  {\bibfnamefont {H.-S.}\ \bibnamefont {Zong}},\ }\href {\doibase
  10.1103/PhysRevD.97.103013} {\bibfield  {journal} {\bibinfo  {journal} {Phys.
  Rev.}\ }\textbf {\bibinfo {volume} {D97}},\ \bibinfo {pages} {103013}
  (\bibinfo {year} {2018}{\natexlab{b}})}\BibitemShut {NoStop}%
\bibitem [{\citenamefont {Nambu}\ and\ \citenamefont
  {Jona-Lasinio}(1961{\natexlab{a}})}]{Nambu:1961a}%
  \BibitemOpen
  \bibfield  {author} {\bibinfo {author} {\bibfnamefont {Y.}~\bibnamefont
  {Nambu}}\ and\ \bibinfo {author} {\bibfnamefont {G.}~\bibnamefont
  {Jona-Lasinio}},\ }\href {\doibase 10.1103/PhysRev.122.345} {\bibfield
  {journal} {\bibinfo  {journal} {Phys. Rev.}\ }\textbf {\bibinfo {volume}
  {122}},\ \bibinfo {pages} {345} (\bibinfo {year}
  {1961}{\natexlab{a}})}\BibitemShut {NoStop}%
\bibitem [{\citenamefont {Nambu}\ and\ \citenamefont
  {Jona-Lasinio}(1961{\natexlab{b}})}]{Nambu:1961b}%
  \BibitemOpen
  \bibfield  {author} {\bibinfo {author} {\bibfnamefont {Y.}~\bibnamefont
  {Nambu}}\ and\ \bibinfo {author} {\bibfnamefont {G.}~\bibnamefont
  {Jona-Lasinio}},\ }\href {\doibase 10.1103/PhysRev.124.246} {\bibfield
  {journal} {\bibinfo  {journal} {Phys. Rev.}\ }\textbf {\bibinfo {volume}
  {124}},\ \bibinfo {pages} {246} (\bibinfo {year}
  {1961}{\natexlab{b}})}\BibitemShut {NoStop}%
\bibitem [{\citenamefont {Klevansky}(1992)}]{Klevansky:1992}%
  \BibitemOpen
  \bibfield  {author} {\bibinfo {author} {\bibfnamefont {S.~P.}\ \bibnamefont
  {Klevansky}},\ }\href {\doibase 10.1103/RevModPhys.64.649} {\bibfield
  {journal} {\bibinfo  {journal} {Rev. Mod. Phys.}\ }\textbf {\bibinfo {volume}
  {64}},\ \bibinfo {pages} {649} (\bibinfo {year} {1992})}\BibitemShut
  {NoStop}%
\bibitem [{\citenamefont {Chubukov}\ \emph {et~al.}(2016)\citenamefont
  {Chubukov}, \citenamefont {Khodas},\ and\ \citenamefont
  {Fernandes}}]{Chubukov:2016}%
  \BibitemOpen
  \bibfield  {author} {\bibinfo {author} {\bibfnamefont {A.~V.}\ \bibnamefont
  {Chubukov}}, \bibinfo {author} {\bibfnamefont {M.}~\bibnamefont {Khodas}}, \
  and\ \bibinfo {author} {\bibfnamefont {R.~M.}\ \bibnamefont {Fernandes}},\
  }\href {\doibase 10.1103/PhysRevX.6.041045} {\bibfield  {journal} {\bibinfo
  {journal} {Phys. Rev. X}\ }\textbf {\bibinfo {volume} {6}},\ \bibinfo {pages}
  {041045} (\bibinfo {year} {2016})}\BibitemShut {NoStop}%
\bibitem [{\citenamefont {Honerkamp}(2008)}]{Honerkamp:2008}%
  \BibitemOpen
  \bibfield  {author} {\bibinfo {author} {\bibfnamefont {C.}~\bibnamefont
  {Honerkamp}},\ }\href {\doibase 10.1103/PhysRevLett.100.146404} {\bibfield
  {journal} {\bibinfo  {journal} {Phys. Rev. Lett.}\ }\textbf {\bibinfo
  {volume} {100}},\ \bibinfo {pages} {146404} (\bibinfo {year}
  {2008})}\BibitemShut {NoStop}%
\bibitem [{\citenamefont {Roy}\ and\ \citenamefont {Foster}(2018)}]{Roy:2018}%
  \BibitemOpen
  \bibfield  {author} {\bibinfo {author} {\bibfnamefont {B.}~\bibnamefont
  {Roy}}\ and\ \bibinfo {author} {\bibfnamefont {M.~S.}\ \bibnamefont
  {Foster}},\ }\href {\doibase 10.1103/PhysRevX.8.011049} {\bibfield  {journal}
  {\bibinfo  {journal} {Phys. Rev. X}\ }\textbf {\bibinfo {volume} {8}},\
  \bibinfo {pages} {011049} (\bibinfo {year} {2018})}\BibitemShut {NoStop}%
\bibitem [{\citenamefont {Tang}\ \emph {et~al.}(2018)\citenamefont {Tang},
  \citenamefont {Leaw}, \citenamefont {Rodrigues}, \citenamefont {Herbut},
  \citenamefont {Sengupta}, \citenamefont {Assaad},\ and\ \citenamefont
  {Adam}}]{Tang:2018}%
  \BibitemOpen
  \bibfield  {author} {\bibinfo {author} {\bibfnamefont {H.-K.}\ \bibnamefont
  {Tang}}, \bibinfo {author} {\bibfnamefont {J.~N.}\ \bibnamefont {Leaw}},
  \bibinfo {author} {\bibfnamefont {J.~N.~B.}\ \bibnamefont {Rodrigues}},
  \bibinfo {author} {\bibfnamefont {I.~F.}\ \bibnamefont {Herbut}}, \bibinfo
  {author} {\bibfnamefont {P.}~\bibnamefont {Sengupta}}, \bibinfo {author}
  {\bibfnamefont {F.~F.}\ \bibnamefont {Assaad}}, \ and\ \bibinfo {author}
  {\bibfnamefont {S.}~\bibnamefont {Adam}},\ }\href {\doibase
  10.1126/science.aao2934} {\bibfield  {journal} {\bibinfo  {journal}
  {Science}\ }\textbf {\bibinfo {volume} {361}},\ \bibinfo {pages} {570}
  (\bibinfo {year} {2018})}\BibitemShut {NoStop}%
\bibitem [{\citenamefont {Cui}\ \emph {et~al.}(2015)\citenamefont {Cui},
  \citenamefont {Hou}, \citenamefont {Shi}, \citenamefont {Wang},\ and\
  \citenamefont {Zong}}]{Cui:2015}%
  \BibitemOpen
  \bibfield  {author} {\bibinfo {author} {\bibfnamefont {Z.-F.}\ \bibnamefont
  {Cui}}, \bibinfo {author} {\bibfnamefont {F.-Y.}\ \bibnamefont {Hou}},
  \bibinfo {author} {\bibfnamefont {Y.-M.}\ \bibnamefont {Shi}}, \bibinfo
  {author} {\bibfnamefont {Y.-L.}\ \bibnamefont {Wang}}, \ and\ \bibinfo
  {author} {\bibfnamefont {H.-S.}\ \bibnamefont {Zong}},\ }\href {\doibase
  10.1016/j.aop.2015.03.025} {\bibfield  {journal} {\bibinfo  {journal} {Annals
  Phys.}\ }\textbf {\bibinfo {volume} {358}},\ \bibinfo {pages} {172} (\bibinfo
  {year} {2015})}\BibitemShut {NoStop}%
\bibitem [{\citenamefont {He}\ \emph {et~al.}(2005)\citenamefont {He},
  \citenamefont {Jin},\ and\ \citenamefont {Zhuang}}]{He:2005}%
  \BibitemOpen
  \bibfield  {author} {\bibinfo {author} {\bibfnamefont {L.}~\bibnamefont
  {He}}, \bibinfo {author} {\bibfnamefont {M.}~\bibnamefont {Jin}}, \ and\
  \bibinfo {author} {\bibfnamefont {P.}~\bibnamefont {Zhuang}},\ }\href
  {\doibase 10.1103/PhysRevD.71.116001} {\bibfield  {journal} {\bibinfo
  {journal} {Phys. Rev. D}\ }\textbf {\bibinfo {volume} {71}},\ \bibinfo
  {pages} {116001} (\bibinfo {year} {2005})}\BibitemShut {NoStop}%
\bibitem [{\citenamefont {Borsanyi}\ \emph {et~al.}(2010)\citenamefont
  {Borsanyi}, \citenamefont {Fodor}, \citenamefont {Hoelbling}, \citenamefont
  {Katz}, \citenamefont {Krieg}, \citenamefont {Ratti},\ and\ \citenamefont
  {Szabo}}]{Borsanyi:2010}%
  \BibitemOpen
  \bibfield  {author} {\bibinfo {author} {\bibfnamefont {S.}~\bibnamefont
  {Borsanyi}}, \bibinfo {author} {\bibfnamefont {Z.}~\bibnamefont {Fodor}},
  \bibinfo {author} {\bibfnamefont {C.}~\bibnamefont {Hoelbling}}, \bibinfo
  {author} {\bibfnamefont {S.~D.}\ \bibnamefont {Katz}}, \bibinfo {author}
  {\bibfnamefont {S.}~\bibnamefont {Krieg}}, \bibinfo {author} {\bibfnamefont
  {C.}~\bibnamefont {Ratti}}, \ and\ \bibinfo {author} {\bibfnamefont {K.~K.}\
  \bibnamefont {Szabo}} (\bibinfo {collaboration} {Wuppertal-Budapest}),\
  }\href {\doibase 10.1007/JHEP09(2010)073} {\bibfield  {journal} {\bibinfo
  {journal} {JHEP}\ }\textbf {\bibinfo {volume} {09}},\ \bibinfo {pages} {073}
  (\bibinfo {year} {2010})}\BibitemShut {NoStop}%
\bibitem [{\citenamefont {Aggarwal}\ \emph {et~al.}(2010)\citenamefont
  {Aggarwal} \emph {et~al.}}]{Aggarwal:2010}%
  \BibitemOpen
  \bibfield  {author} {\bibinfo {author} {\bibfnamefont {M.~M.}\ \bibnamefont
  {Aggarwal}} \emph {et~al.} (\bibinfo {collaboration} {STAR Collaboration}),\
  }\href {\doibase 10.1103/PhysRevLett.105.022302} {\bibfield  {journal}
  {\bibinfo  {journal} {Phys. Rev. Lett.}\ }\textbf {\bibinfo {volume} {105}},\
  \bibinfo {pages} {022302} (\bibinfo {year} {2010})}\BibitemShut {NoStop}%
\bibitem [{\citenamefont {Adamczyk}\ \emph {et~al.}(2014)\citenamefont
  {Adamczyk} \emph {et~al.}}]{Adamczyk:2014}%
  \BibitemOpen
  \bibfield  {author} {\bibinfo {author} {\bibfnamefont {L.}~\bibnamefont
  {Adamczyk}} \emph {et~al.} (\bibinfo {collaboration} {STAR Collaboration}),\
  }\href {\doibase 10.1103/PhysRevLett.112.032302} {\bibfield  {journal}
  {\bibinfo  {journal} {Phys. Rev. Lett.}\ }\textbf {\bibinfo {volume} {112}},\
  \bibinfo {pages} {032302} (\bibinfo {year} {2014})}\BibitemShut {NoStop}%
\bibitem [{\citenamefont {Luo}(2016)}]{Luo:2015}%
  \BibitemOpen
  \bibfield  {author} {\bibinfo {author} {\bibfnamefont {X.}~\bibnamefont
  {Luo}},\ }\href {\doibase 10.1016/j.nuclphysa.2016.03.025} {\bibfield
  {journal} {\bibinfo  {journal} {Nucl. Phys.}\ }\textbf {\bibinfo {volume}
  {A956}},\ \bibinfo {pages} {75} (\bibinfo {year} {2016})}\BibitemShut
  {NoStop}%
\bibitem [{\citenamefont {Luo}\ and\ \citenamefont {Xu}(2017)}]{Luo:2017}%
  \BibitemOpen
  \bibfield  {author} {\bibinfo {author} {\bibfnamefont {X.}~\bibnamefont
  {Luo}}\ and\ \bibinfo {author} {\bibfnamefont {N.}~\bibnamefont {Xu}},\
  }\href {\doibase 10.1007/s41365-017-0257-0} {\bibfield  {journal} {\bibinfo
  {journal} {Nucl. Sci. Tech.}\ }\textbf {\bibinfo {volume} {28}},\ \bibinfo
  {pages} {112} (\bibinfo {year} {2017})}\BibitemShut {NoStop}%
\bibitem [{\citenamefont {Adamczyk}\ \emph {et~al.}(2018)\citenamefont
  {Adamczyk} \emph {et~al.}}]{Adamczyk:2017}%
  \BibitemOpen
  \bibfield  {author} {\bibinfo {author} {\bibfnamefont {L.}~\bibnamefont
  {Adamczyk}} \emph {et~al.} (\bibinfo {collaboration} {STAR}),\ }\href
  {\doibase 10.1016/j.physletb.2018.07.066} {\bibfield  {journal} {\bibinfo
  {journal} {Phys. Lett.}\ }\textbf {\bibinfo {volume} {B785}},\ \bibinfo
  {pages} {551} (\bibinfo {year} {2018})}\BibitemShut {NoStop}%
\bibitem [{\citenamefont {Gupta}\ \emph {et~al.}(2011)\citenamefont {Gupta},
  \citenamefont {Luo}, \citenamefont {Mohanty}, \citenamefont {Ritter},\ and\
  \citenamefont {Xu}}]{Gupta:2011}%
  \BibitemOpen
  \bibfield  {author} {\bibinfo {author} {\bibfnamefont {S.}~\bibnamefont
  {Gupta}}, \bibinfo {author} {\bibfnamefont {X.}~\bibnamefont {Luo}}, \bibinfo
  {author} {\bibfnamefont {B.}~\bibnamefont {Mohanty}}, \bibinfo {author}
  {\bibfnamefont {H.~G.}\ \bibnamefont {Ritter}}, \ and\ \bibinfo {author}
  {\bibfnamefont {N.}~\bibnamefont {Xu}},\ }\href {\doibase
  10.1126/science.1204621} {\bibfield  {journal} {\bibinfo  {journal}
  {Science}\ }\textbf {\bibinfo {volume} {332}},\ \bibinfo {pages} {1525}
  (\bibinfo {year} {2011})}\BibitemShut {NoStop}%
\bibitem [{\citenamefont {Abbott}\ \emph {et~al.}(2017)\citenamefont {Abbott}
  \emph {et~al.}}]{Abbott:2017prl}%
  \BibitemOpen
  \bibfield  {author} {\bibinfo {author} {\bibfnamefont {B.}~\bibnamefont
  {Abbott}} \emph {et~al.} (\bibinfo {collaboration} {LIGO Scientific,
  Virgo}),\ }\href {\doibase 10.1103/PhysRevLett.119.161101} {\bibfield
  {journal} {\bibinfo  {journal} {Phys. Rev. Lett.}\ }\textbf {\bibinfo
  {volume} {119}},\ \bibinfo {pages} {161101} (\bibinfo {year}
  {2017})}\BibitemShut {NoStop}%
\bibitem [{\citenamefont {Antoniadis}\ \emph {et~al.}(2013)\citenamefont
  {Antoniadis} \emph {et~al.}}]{Antoniadis:2013}%
  \BibitemOpen
  \bibfield  {author} {\bibinfo {author} {\bibfnamefont {J.}~\bibnamefont
  {Antoniadis}} \emph {et~al.},\ }\href {\doibase 10.1126/science.1233232}
  {\bibfield  {journal} {\bibinfo  {journal} {Science}\ }\textbf {\bibinfo
  {volume} {340}},\ \bibinfo {pages} {6131} (\bibinfo {year}
  {2013})}\BibitemShut {NoStop}%
\end{thebibliography}%
\end{document}